\documentclass[twocolumn,prl,preprintnumbers,superscriptaddress,showpacs]{revtex4}

\usepackage{graphicx}
\usepackage{amsmath,amsfonts,amssymb,amsxtra}
\usepackage{units,formula}

\makeatletter
\DeclareRobustCommand{\chemical}[1]{%
  {\(\m@th
   \edef\resetfontdimens{\noexpand\)%
       \fontdimen16\textfont2=\the\fontdimen16\textfont2
       \fontdimen17\textfont2=\the\fontdimen17\textfont2\relax}%
   \fontdimen16\textfont2=2.7pt \fontdimen17\textfont2=2.7pt
   \mathrm{#1}%
   \resetfontdimens}}
\DeclareRobustCommand{\bchemical}[1]{%
  {\(\m@th
   \edef\resetfontdimens{\noexpand\)%
       \fontdimen16\textfont2=\the\fontdimen16\textfont2
       \fontdimen17\textfont2=\the\fontdimen17\textfont2\relax}%
   \fontdimen16\textfont2=2.7pt \fontdimen17\textfont2=2.7pt
   \mathbf{#1}%
   \resetfontdimens}}
\makeatother

\newcommand{\half}{\chemical{La_{0.5}Sr_{1.5}MnO_4}}
\newcommand{\hole}{\chemical{La_{0.42}Sr_{1.58}MnO_4}}

\newcommand{\TN}{T$_\text{N}$}

\newcommand{\Q}{\bm Q}
\newcommand{\q}{\bm q}

\newcommand{\Mnd}{Mn$^{3+}$}
\newcommand{\Mnv}{Mn$^{4+}$}
\begin{document}

\textheight 24.4 true cm

\title{ Evidence for charge orbital and spin stripe order in an overdoped manganite }
\author{H. Ulbrich}
\affiliation{II.~Physikalisches Institut, Universit\"at zu K\"oln,
Z\"ulpicher Str.~77, D-50937 K\"oln, Germany}

\author{D. Senff} \affiliation{II.~Physikalisches Institut, Universit\"at zu K\"oln, Z\"ulpicher Str.~77, D-50937 K\"oln, Germany}

\author{P. Steffens}
\affiliation{Institut Laue-Langevin, BP 156, 38042 Grenoble Cedex
9, France}

\author{O.J. Schumann} \affiliation{II.~Physikalisches Institut, Universit\"at zu K\"oln, Z\"ulpicher Str.~77, D-50937 K\"oln, Germany}

\author{Y. Sidis}
\affiliation{Laboratoire L\'eon Brillouin, C.E.A./C.N.R.S.,
F-91191 Gif-sur-Yvette Cedex, France}

\author{P. Reutler} \affiliation{Laboratoire de Chimie des Solides, LPCES - ICMMO, Universit\'e de Paris Sud XI, 91405 Orsay, France}

\author{A. Revcolevschi} \affiliation{Laboratoire de Chimie des Solides, LPCES - ICMMO, Universit\'e de Paris Sud XI, 91405 Orsay, France}

\author{M. Braden}
\email{braden@ph2.uni-koeln.de} \affiliation{II.~Physikalisches
Institut, Universit\"at zu K\"oln, Z\"ulpicher Str.~77, D-50937
K\"oln, Germany}

\date{\today, \textbf{preprint}}

\pacs{75.20.-m, 71.10.-w, 75.47.Lx, 75.50.Ee, 75.25.+z}

\date{\today}
\begin{abstract}

Overdoped La$_{0.42}$Sr$_{1.58}$MnO$_4$ exhibits a complex ordering of charges, orbitals and spins. Single-crystal neutron diffraction
experiments reveal three incommensurate and one commensurate order parameters to be tightly coupled. The position and the shape of the distinct
superstructure scattering points to a stripe arrangement in which ferromagnetic zigzag chains are disrupted by additional Mn$^{4+}$ stripes. The
observation of remarkably strong higher harmonics scattering for the orbital ordering corroborates this picture.

\end{abstract}

\maketitle

In the stripe phases first observed in cuprates \cite{Tranq1}
additional charges segregate into stripes so that an
antiferromagnetic order in between is conserved. The stripe
scenario which can be considered as a soliton lattice is of
general relevance in the physics of doped Mott insulators with
similar stripe patterns being established in nickelates and in
cobaltates \cite{Tranq1,Tranq2,Kaji,Cwik}. The role of the orbital
degree of freedom in manganites implies an additional aspect to
stripe ordering as the magnetic interaction directly follows the
orbital arrangement. Evidence of incommensurate ordering of
charges and orbitals has also been observed for various manganites
at higher doping levels, i.e. overdoped with respect to half
doping with stable charge and orbital ordering (COO) but a clear
picture of these phases, in particular the coupling to the
magnetism, is still missing in spite of strong efforts
\cite{5,6,7,8,9,10,11,12,13,14,15,16,17,18,19,pissa}.

First evidence for an incommensurate structural ordering in manganites was obtained from electron diffraction experiments which find the
incommensurate modulations to linearly follow the amount of extra charges \cite{5,6,7}. This fact was interpreted as evidence for stripe phases
where the extra \Mnv \ segregate into stripes along the diagonals of the simple perovskite lattice. In the meanwhile, numerous electron, x-ray
and neutron diffraction experiments studied the stripe phases in various overdoped manganites, however, without reaching a clear conclusion
\cite{8,9,10,13,14,18,pissa}. So far it seems impossible to decide between the two proposed models: the Wigner crystal and the bi-stripe
arrangement \cite{5,6,7}. Furthermore, even the intrinsic nature of the structural phases remains matter of controversy. Both a soliton lattice
reflecting the stripe pattern \cite{11,12} and a homogenous charge-density wave \cite{13} have been proposed to explain the incommensurabilities
generating strong controversy.

We have studied the charge, orbital and spin ordering (COSO) in
overdoped La$_{0.42}$Sr$_{1.58}$MnO$_4$ by single-crystal neutron
diffraction. Three incommensurate order parameters associated with
charges, orbitals and Mn$^{3+}$ spins, respectively, and one
commensurate one are tightly coupled with each other. The position
and the shape of the different kinds of diffuse scattering as well
as the observation of higher harmonics indicate a stripe-type
arrangement where stripes of additional charges disrupt the
ferromagnetic zigzag chains.

\begin{figure}
\includegraphics[width=0.48\textwidth]{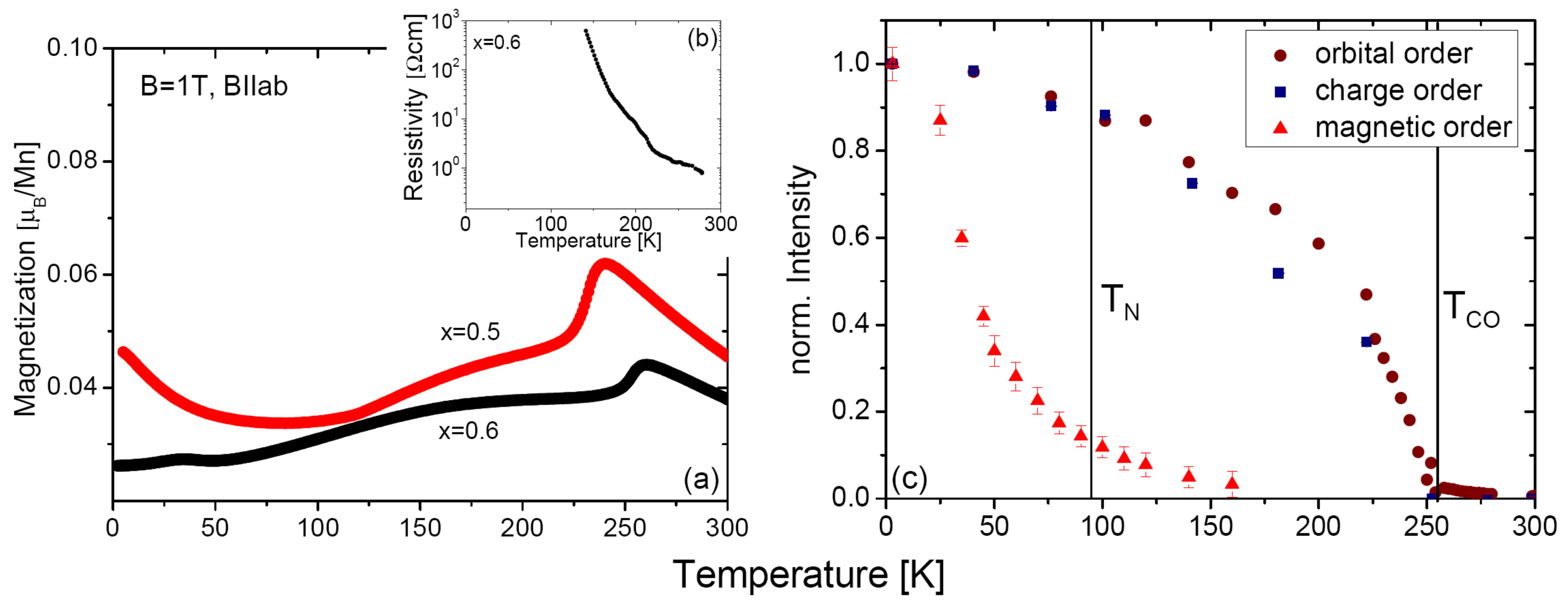}

\caption{Temperature dependence of magnetization as a function of
temperature for a field of 1T applied parallel to the ab-planes
for the half and for the overdoped compound (a). The inset
presents the in-plane electric resistivity for x=0.58 (b).
Characteristic superstructure reflections measured by neutron
diffraction increase in intensity at the transition temperature
into the COO-state and into the AFM-state, respectively (c).}
\label{makr}
\end{figure}

\begin{figure*}
\includegraphics[width=0.9\textwidth]{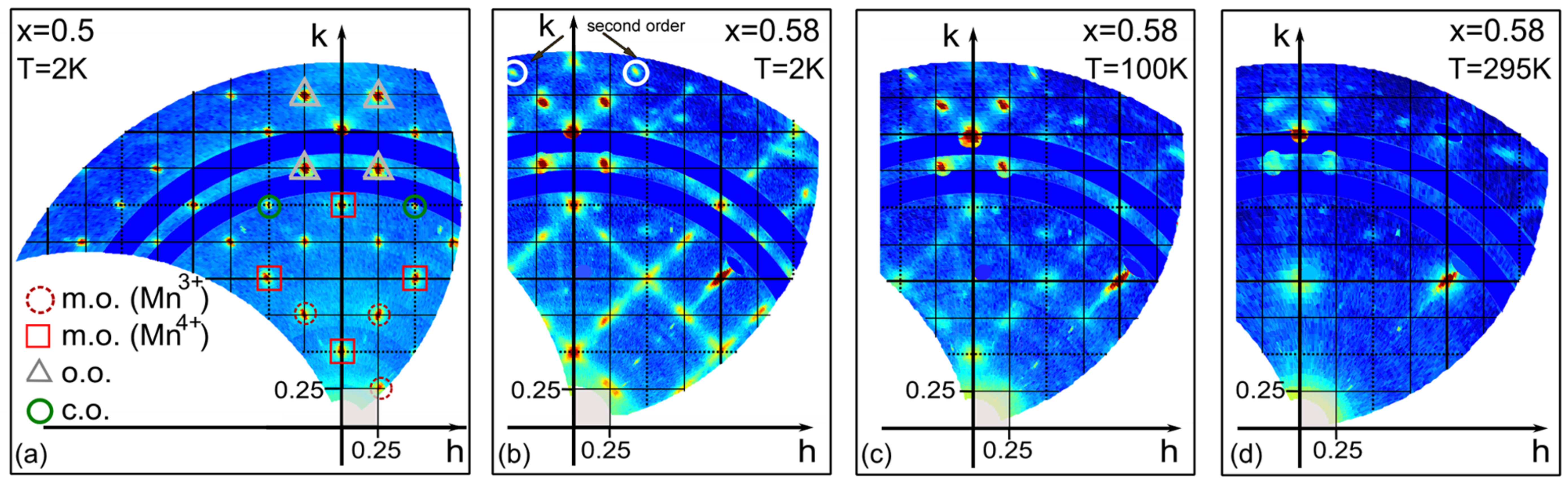}
\caption{Distribution of structural and magnetic scattering in the (h,k,0)-plane for {\half} and \hole. For x=0.5 superstructure reflections
refer to the charge (green circles), orbital (grey triangles) and magnetic ordering ({\Mnd}: dotted circles; {\Mnv} red squares) (a). The
COSO-state for x=0.58 is governed by an incommensurate ordering of charge, orbital and magnetic ordering of {\Mnd} (b). Slightly above the
N\'eel-temperature {\TN} the magnetic superstructure reflections vanish but a diffuse magnetic signal is still visible (c). At room-temperature
diffuse signals exist around the Bragg-reflections {\bf Q}=(0 2 0) and {\bf Q}=(0 1 0) (d). Contaminations from a crystallite and from the Al
sample-holder are covered in blue color.} \label{Flatcone}
\end{figure*}

A large single crystal of {\hole} was grown by the floating-zone technique \cite{26}. {\hole} crystallizes in a tetragonal structure of space
group I4/mmm with room-temperature lattice constants a$_{0}$=3.86 {\AA} and c$_{0}$=12.40 \AA. The magnetization was measured by a SQUID
magnetometer and electric resistivity by a standard four-contact method (Fig. \ref{makr}). The transition into the COO state
(T$_{\text{CO}}$=255K) manifests itself by the appearance of superstructure reflections, as well as by a sharp drop of the magnetization M(T).
The antiferromagnetic ordering, however, is not visible in the macroscopic data most likely due to a strong 2D character. Magnetic
superstructure reflections are detected below \TN $\approx$ 95K  in good agreement with \cite{19}, but the transition is sluggish. The neutron
diffraction experiments were performed at the thermal diffractometer 3T.1 at the Laboratoire L\'eon Brillouin and at the triple-axis
spectrometer IN20 at the Institut Laue Langevin. On 3T.1 we used a neutron energy of 14.7meV and two pyrolytic graphite (PG) filters to suppress
higher order contaminations. On IN20 the flatcone detector with silicon (111) analyzers fixing the final energy to 18.7 meV allowed us to detect
full maps of scattering in an efficient way. In all experiments, the c-axis was set vertical to the diffraction plane.

The COSO phase in \half \ has been characterized in great detail using various scattering methods \cite{18,22,D_FM}. With the flatcone detector
we may easily map the hk0-plane and identify the different types of characteristic scattering. With respect to the symmetry of the tetragonal
high-temperature paramagnetic phase four different order parameters can be separated, see Ref. \onlinecite{D_FM}. Proper charge ordering of the
checkerboard type is related to a mode at {\bf k}$_{\text{co}}$ =$\pm$(0.5,0.5) (referring to the 2-dimensional space and thus neglecting
inter-layer coupling). The orbital ordering causes an additional doubling along one direction and is associated with a mode at \q {\bf
k}$_{\text{oo}}$=$\pm$(0.25,0.25). The magnetism is described by two propagation vectors {\bf k}$_{\text{Mn}^{3+}}$=$\pm$(0.25,-0.25) and {\bf
k}$_{\text{Mn}^{4+}}$=$\pm$(0.5,0.0) referring to the nominal \Mnd \ and \Mnv \ spins, respectively. In the scattering map in Fig. 2(a) one may
thus easily attribute each superstructure scattering with the associated order parameter. Just for the quarter-indexed peaks the identification
is not intrinsic but can nevertheless easily be obtained due to the distinct {\bf Q}-dependence of magnetic and structural scattering.

Fig. \ref{Flatcone}(b) illustrates the same area of reciprocal space at T=2K for the overdoped system \hole. The single-counter scans across the
different scattering types are shown in Fig. 3. The inspection of the two scattering maps immediately shows that the ordering phenomena in the
two samples are qualitatively similar, but instead of the sharp signals at commensurate {\bf \Q}-positions, there are broader features mostly
centered at incommensurate positions in \hole . The direct comparison allows one to identify three underlying ordering schemes. Superstructure
reflections referring to the charge ordering are centered  for example at {\bf Q}=(1.5+\text{$\delta_{\text{co}}$}
1.5+\text{$\delta_{\text{co}}$} 0) with \text{$\delta_{\text{co}}$}=0.080(3). The incommensurability $\delta_{\text{co}}$ does not change with
temperature as scans parallel to the modulation exhibit only the thermal suppression of intensity, see Fig. 3(a). Fig. \ref{Flatcone}(b)
demonstrates that the superstructure reflections associated with the orbital ordering are incommensurate, as well in agreement with previous
x-ray experiments \cite{18}. The orbital satellites are centered closer to the Bragg-reflection, for example {\bf Q}=(0 2 0) in diagonal
direction. At this {\bf \Q} range magnetic scattering can be fully neglected due to the magnetic form factor. The orbital scattering is
incommensurately displaced towards the central Bragg peak, {\bf k}$_{\text{oo}}$=(0.25-$\delta_{\text{oo}}$,0.25-$\delta_{\text{oo}}$) and again
there is no temperature dependency, see Fig. 3(b). The incommensurability of the orbital satellites is within the error bars exactly  half of
that of  the charge ordering (\text{\text{$\delta_{\text{oo}}$}=0.039(2)=$\frac{1}{2}\delta_{\text{co}}$}). These observations reveal the tight
coupling between charge and orbital ordering in \hole , and they perfectly agree with previous diffraction studies on overdoped perovskite and
bi-layer manganites using various diffraction methods \cite{5,6,7,8,9,10,13,14,18}. Most interestingly we clearly observe remarkably strong
second-order harmonics of the orbital signal at (2,0,0)$\pm$2{\bf k}$_{\text{oo}}$, see Fig. 2(b) and (c).

\begin{figure}
\includegraphics[width=0.45\textwidth]{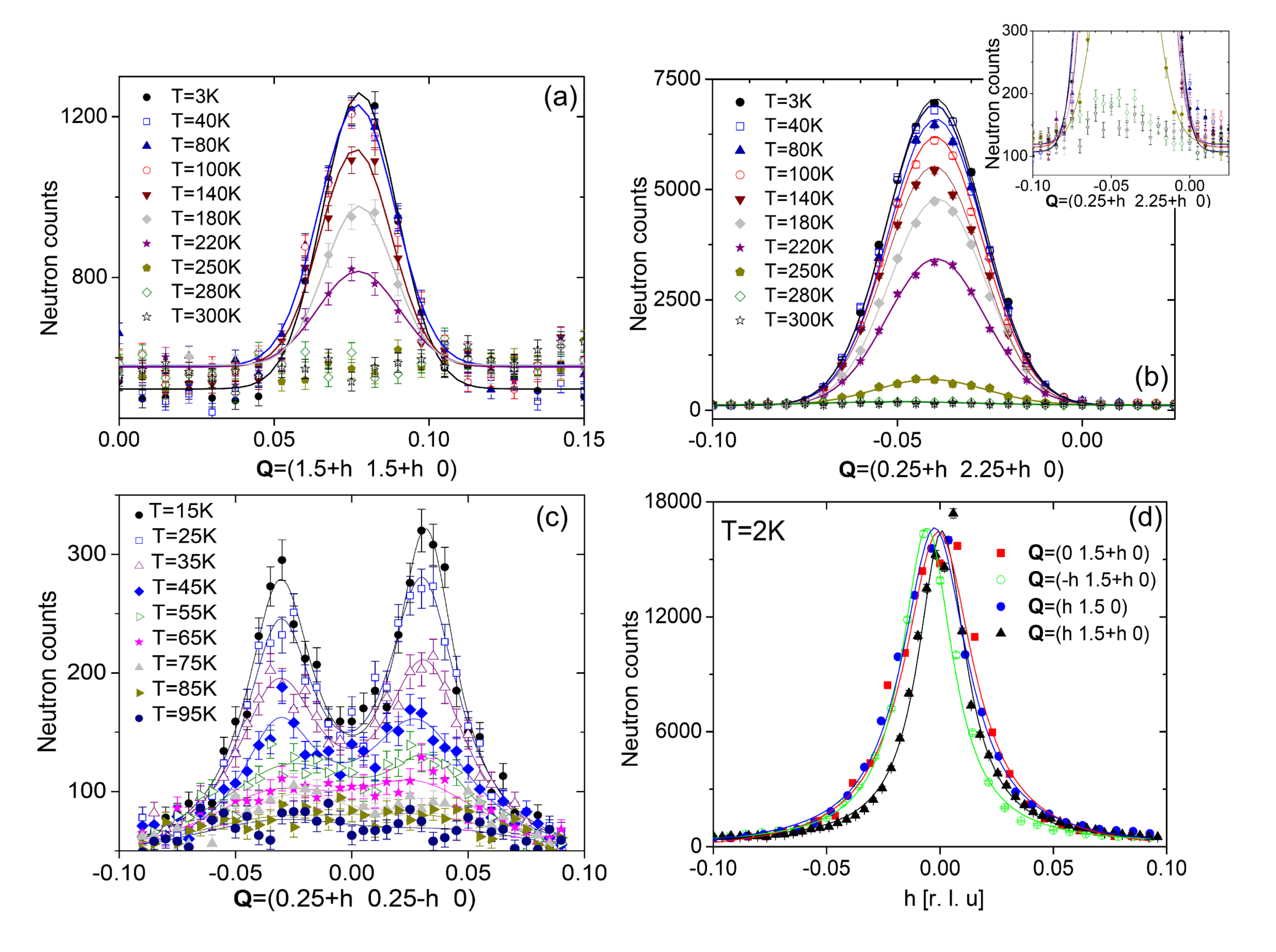}
\caption{Elastic scans across the positions of the different
superstructure reflections. Satellites referring to the charge
ordering (a) are displaced twice as much as those of orbital
ordering (b) and of {\Mnd} spin ordering  (c), respectively.
Reflections related to the magnetic ordering of {\Mnv} appear at
commensurate positions (d).} \label{single}
\end{figure}

Inspection of Fig. \ref{Flatcone}(b) indicates that the quarter-indexed scattering associated with \Mnd -spin ordering at half doping is split
into two contributions in \hole . The incommensurate modulation of the \Mnd -spin order is transverse with respect to the {\Mnd} order in x=0.5
(but note it is parallel to the chains) {\bf k}$_{\text{Mn}^{3+}}$=(0.25-$\delta_{\text{Mn}^{3+}}$,0.25+$\delta_{\text{Mn}^{3+}}$), see Fig.
3(c). Scans in longitudinal direction yield a peak centered at the commensurate position due to the overlap of the two incommensurate
contributions. As in Reference \onlinecite{19} only longitudinal scans were performed, the incommensurate character of the magnetic order was
overlooked. The displacement $\delta_{\text{Mn}^{3+}}$ exhibits no temperature dependence. From several magnetic superstructure reflections we
determine the average incommensurability $\delta_{\text{Mn}^{3+}}$ which perfectly agrees with the incommensurability of the orbital ordering:
\text{\text{$\delta_{\text{Mn}^{3+}}$}=0.037(2)=$\delta_{\text{oo}}$}=$\frac{1}{2}\delta_{\text{co}}$. These three independent order parameters
are all incommensurate and tightly coupled to each other.

The magnetic scattering associated with the ordering of {\Mnv} spins are centered at the same positions as those in \half (see Fig.
\ref{Flatcone}), but the scattering in \hole \ exhibits diffuse tails along [110] direction. The commensurate ordering of {\Mnv} spins appears
astonishing as magnetic and orbital degrees of freedom are typically strongly coupled. Therefore additional scans were performed across the
commensurate position {\bf Q}$_{\text{Mn}^{4+}}$=$\pm$(0.0,1.5) in horizontal, vertical and diagonal directions. None of them indicates a
sizeable incommensurability. Therefore we may safely exclude, that the \Mnv -spin scattering exhibits an incommensurate shift comparable to
those of the other three order parameters.

\begin{figure}
\includegraphics[width=0.36\textwidth]{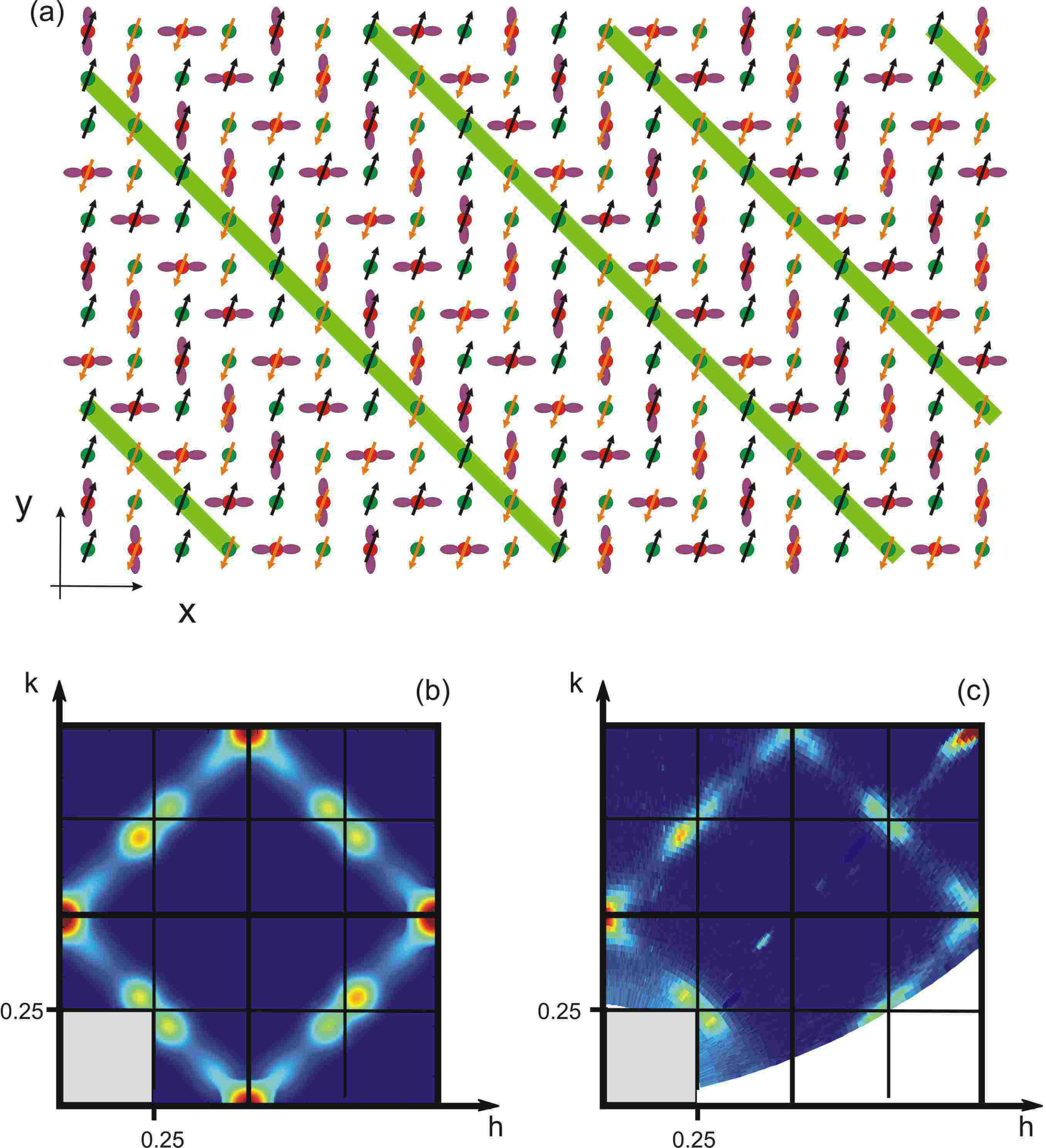}
\caption{Sketch of charge, orbital and spin order for \hole \ (a).
Red circles represent {\Mnd} and green circles {\Mnv}. A single
domain of zigzag chains propagating in [110] direction is shown.
The excess lines of {\Mnv} are displayed on a green background.
The Fourier transformations superposed for all domains (b) agree
with the experimental map (c). Note, that the intensity at the
position Q=(1 1 0) in (c) is a
 Bragg peak.} \label{FT}
\end{figure}
Because of an excess of {\Mnv} ions an ideal checkerboard order of charges is not possible. The main question thus concerns the arrangement of
these extra \Mnv . Since all three observed incommensurabilities appear along the diagonal, these extra \Mnv \ arrange in stripes along the
diagonals as assumed already from the orbital reflections detected previously. This arrangement resembles the stripe phases observed in
cuprates, nickelates and cobaltates \cite{Tranq1,Tranq2,Kaji, Cwik} and corresponds to a soliton lattice. We may further deduce from the orbital
incommensurate modulation, which is longitudinal in nature, that the disruption of the orbital order occurs along the zigzag-chains, or in other
words that the \Mnv -stripes align perpendicular to the chains. The magnetic superstructure reflections are in perfect agreement with this. The
magnetic propagation vector {\bf k}$_{\text{Mn}^{3+}}$ points exactly perpendicular to the zigzag chains at half doping, so that the arrangement
of the {\Mnv} stripes perpendicular to the zigzag-chains results in a transverse modulation of the corresponding magnetic order. The resulting
real-space model of this configuration is shown in Fig. 4.

It may appear astonishing that in spite of the stripe arrangement the magnetic scattering  associated with the \Mnv -spins at {\bf
k}$_{\text{Mn}^{4+}}$ remains commensurate, however, this fact is easily explained. The magnetic structure in the zigzag fragments can be
constructed by following the strong ferromagnetic interaction across the \Mnd \ orbitals. However, this does not fix the magnetic coupling in
the double line of \Mnv . Ferromagnetic \Mnv -spin pairs in these double lines may align either horizontally or vertically (the vertical
solution is shown in Fig. 4(a), but all these pairs must align in the same sense. The two possibilities, however, do not create distinct
magnetic symmetries but only distinct domains; the symmetry of this magnetic structure is monoclinic generating thus four domain orientations
with respect to the tetragonal high-temperature phase. This magnetic structure causes the commensurate {\bf k}$_{\text{Mn}^{4+}}$ intensities,
as the \Mnv -spins sum up along the vertical bond direction and alternate in the horizontal direction resulting in magnetic scattering at {\bf
\Q} =(0.5,0) for the domain chosen in Fig. 4.

In order to analyze the position and the shape of the different scattering contributions we performed two-dimensional Fourier-transformations of
large supercells (typically 380 $\cdot$ 380 lattices) in which the charge, orbital and spin order were implemented schematically. If we
calculate the average incommensurability of orbital, of magnetic ({\Mnd}), and half that of charge ordering, we obtain $\bar\delta$=0.039(2)
which via the scaling relation $\epsilon (x)$ indicates a Sr content x=0.58.  This value perfectly agrees with a single-crystal structure
determination by x-ray diffraction, but is slightly below the nominal composition of the starting rod in the crystal growth \cite{26}. The
concentration x=0.58 can not be realized by an adjustment of stripes with equal distances, but by a mixture of 33\% blocks being arranged
according to x=0.6 and 66\%  blocks with x=0.57. The entire modulation consists thus of an alternation of blocks with seven lattice-spacings
length and  with five lattice-spacings length with a ratio of 2:1 [see Fig. \ref{FT}(a)]. The structural arrangement consists thus of a soliton
lattice in which only the average soliton distance is defined. The supercells of about 380$\cdot$380 lattices are larger than the correlation
lengths of the ordering in these manganites (and also larger than the neutron coherence length) so that finite-size effects can be neglected.
The Fourier transformation fully explains the appearance of all the scattering observed, when the superposition of the four domains is taken
into account \footnote{Note, that there seems to be a slight imbalance in the domain occupation in the experimental map, and that we reduced the
moment of the excess Mn$^{4+}$.}. The experimental maps indicate that the magnetic scattering is not sharp but exhibits diffusive tails along
the diagonal concerning both the \Mnv \ and the \Mnd \ contributions; this effect is perfectly reproduced by the Fourier transformation as well.
Magnetic order in the direction parallel to the stripes is perfect on a large length scale, but in the direction perpendicular to the stripes,
magnetic order is regularly perturbed by the soliton generating diffuse tails. Such an anisotropic broadening was shown to yield direct evidence
for a stripe arrangement \cite{28}. The perfect agreement between the simulated and the experimental scattering maps including the diffuse tails
gives strong evidence for the stripe arrangement and excludes a homogeneous spin and charge density wave modulation. This conclusion is further
corroborated by the observation of the remarkably strong second-order harmonics of the orbital superstructure scattering, see Fig. 2(b), which
exclude a simple sinusoidal density wave. The squaring up of a density wave is usually associated with third-order harmonics, but for the
orbital pattern there is no squaring up but a strong pair or triple of opposite orbital phases (associated with a single "zig" in the chains) is
surrounded by suppression of orbital order in the Mn$^{4+}$ rows. This arrangement perfectly agrees with a strong second order harmonics.
Evidence for a magnetic soliton arrangement in overdoped manganites was recently obtained by NMR experiments \cite{17}.

Above the onset of three-dimensional magnetic ordering at T$_\text{{N}}\approx\text{95}$K we still observe a diffuse signal around the quarter-
and half-indexed positions [see Fig.\ref{Flatcone}(c)],
but these signals rapidly lose intensity upon heating and a ferromagnetic signal increases in intensity [see Fig. 2(c)]. In the scattering map
taken above the CO-state (T$>$T$_\text{{CO}}$) the CE-type magnetic scattering is fully lost as it is also seen in the single-counter scans. The
CE-type magnetic correlations get replaced by in-plane ferromagnetic correlations centered around {\bf Q}=(0 1 0) [Fig. \ref{Flatcone}(d)]
reflecting the increase of the macroscopic magnetization, see {Fig. \ref{makr}(a). The competition between the ferromagnetic short-range
magnetic correlations and the CE-type low-temperature magnetic order reported for \half \ \cite{23,D_FM} thus persists in the overdoped
concentration range.

The orbital and charge order scattering persists above T$_\text{{N}}$ without any indication for a sizeable modification upon passing the
magnetic transition. Although the charge scattering apparently fully disappears at the charge and orbital order transition, a sizeable amount of
orbital scattering persists at T=295K, i.e. largely above the phase transition. This scattering is diffusive but it remains clearly
incommensurate. The charge and orbital disordered state thus already exhibits the instability against incommensurate orbital ordering with the
position fixed through the amount of doping. This incommensurate orbital instability cannot be directly caused by the magnetic correlations, as
the orbital instability coexists with dominant ferromagnetic correlations \cite{29}.

In conclusion the position and the shape of the distinct types of
superstructure scattering combined with the strong higher
harmonics unambiguously indicate a stripe-like ordering in
overdoped manganites.

This work was supported by the Deutsche Forschungsgemeinschaft
through the Sonderforschungsbereich 608.




\end{document}